\documentclass[12pt]{iopart}
\usepackage{graphicx}
\usepackage{epsfig}

\begin{document}

\title[The L\"{o}wdin orthogonalization and  magnetoelectric coupling]{The L\"{o}wdin orthogonalization and  magnetoelectric coupling for noncentrosymmetric ions}

%\title[The L\"{o}wdin orthogonalization and  quasi-atoms ]{The L\"{o}wdin orthogonalization and quasi-atoms with a novel magnetoelectric coupling in crystals}

\author{A.S. Moskvin}

\address{Department of Theoretical Physics, Ural Federal University, 620083 Ekaterinburg,  Russia}
\ead{alexandr.moskvin@usu.ru}
\begin{abstract}
%Making use of the L\"{o}wdin orthogonalization for one-electron atomic $\varphi_{fnlm}({\bf r})$ wave functions we arrive at a basis set of orthogonalized $\psi_{fnlm}({\bf r})$ orbitals which formally preserve symmetric properties of the bare $\varphi_{fnlm}({\bf r})$ orbitals. Instead of many-electron atomic configurations composed of non-orthogonalized $nlm$-orbitals we arrive at quasi-atoms composed of the orthogonalized $nlm$-counterparts. 
The L\"{o}wdin orthogonalization procedure being the well-known technique, particularly in quantum chemistry, however, gives rise to novel effects missed in earlier studies. 
Making use of the technique of irreducible tensorial operators we have developed a regular procedure for account of the orthogonalization effects.
 For illustration we address the emergence of a specific magnetoelectric coupling for noncentrosymmetric 3d or 4f ions. 
\end{abstract}
%\submitto{\SUST}
\maketitle

%31.15.-p Calculations and mathematical techniques in atomic and molecular physics (excluding electron correlation calculations)
%71.10.-w Theories and models of many-electron systems
%Multiferroics, 75.85.+t
%Magnetoelectric effects, 75.85.+t

\section{Introduction} 

More than 50 years ago P.-O. L\"{o}wdin\,\cite{Lowdin} suggested a regular procedure for the orthogonalization of the atomic functions localized at different sites. The  orthogonalization problem was in the focus of the electron structure calculations in 60-70th though later on it was undeservedly forgotten in the shade of the a so-called "{\it ab-initio}" calculations. However, many interesting points have been missed being overboard the  orthogonalization procedure. Below in the paper we do generalize the L\"{o}wdin technique for many-electron atoms and demonstrate that the irreducible tensorial operator method which is well known in theory of free atoms can be succesfully applied to "orthogonalized" electrons. As an urgent issue we address the orthogonalization contribution to the orbital magnetoelectric coupling.

The paper is organized as follows. In Sec.II we demonstrate the conservation of the effective rotation symmetry under the L\"{o}wdin orthogonalization technique. In Sec.III and IV we address some novel properties of different operators acting on the orthogonalized basis set.
In Sec.V and VI we do calculate the electric dipole moment of the noncentrosymmetric quasi-atoms   and the orthogonalization contribution to the orbital magnetoelectric coupling.

\section{L\"{o}wdin orthogonalization procedure}
Making use of the one-electron  $\varphi_{fnlm}({\bf r})$ wave functions of a free atom as a basis set for description of the electronic structure in crystals is restricted because of their nonorthogonality for different sites/atoms. One of the more practicable techniques for solution the problem was suggested by L\"{o}wdin\,\cite{Lowdin}. 
Let start with a set of standard one-electron functions $\varphi_{\kappa}({\bf r})=\varphi_{fnlm}({\bf r})$, where $f$ labels a site, and introduce a Hermitian overlap matrix  ${\hat \Delta}={\hat {\bf 1}}+{\hat {\bf S}}$
as follows:
\begin{equation}
{\hat \Delta}_{\kappa\kappa^{'}}=\langle\varphi_{\kappa}({\bf r})|\varphi_{\kappa^{'}}({\bf r})\rangle =\delta_{\kappa\kappa^{'}}+S_{\kappa\kappa^{'}}\, .
\end{equation}
Linear transformation $\varphi \rightarrow \psi =\varphi {\hat {\bf A}}$
\begin{equation}
\psi_{\kappa}({\bf r})=	\sum_{\kappa^{'}}\varphi_{\kappa^{'}}({\bf r})A_{\kappa^{'}\kappa}\, ,
\end{equation}
yields a set of orthogonalized functions $\psi_{\kappa}({\bf r})$:
$$
\langle\psi_{\kappa}({\bf r})|\psi_{\kappa^{'}}({\bf r})\rangle =\delta_{\kappa\kappa^{'}}
$$
Easy to see that matrix elements ${\hat {\bf A}}$ obey a matrix equation
\begin{equation}
{\hat {\bf A}}^{\dagger}{\hat \Delta}{\hat {\bf A}}={\hat {\bf 1}}	\, ,
\end{equation}
whose solution can be represented as follows
\begin{equation}
{\hat {\bf A}}={\hat \Delta}^{-1/2}{\hat {\bf B}}	\, ,
\end{equation}
where ${\hat {\bf B}}^{\dagger}{\hat {\bf B}}={\hat {\bf 1}}$ given the existence of the    ${\hat \Delta}^{-1/2}=({\hat {\bf 1}}+{\hat {\bf S}})^{-1/2}$ matrix. Obviously the choice of the unitary ${\hat {\bf B}}$ matrix, hence the certain orthogonalization  procedure, is ambiguous one. The most physically clear and practicable procedure of a so-called symmetric orthogonalization when ${\hat {\bf B}}={\hat {\bf 1}}$ и ${\hat {\bf A}}={\hat \Delta}^{-1/2}=({\hat {\bf 1}}+{\hat {\bf S}})^{-1/2}$ was suggested by L\"{o}wdin\,\cite{Lowdin}. Given small overlap when $\sum_{\kappa^{'}}|S_{\kappa^{'}\kappa}|\leq q <1$ for all $\kappa$, the series 
\begin{equation}
({\hat {\bf 1}}+{\hat {\bf S}})^{-1/2}={\hat {\bf 1}}-\frac{1}{2}{\hat {\bf S}}+\frac{3}{8}{\hat {\bf S}}^2+\ldots
\end{equation}
will converge, that enables to  represent the orthogonalized functions  $\psi_{\kappa}$ as follows
$$
\psi_{\kappa}({\bf r})=\sum_{\kappa^{'}}\left((1+\hat S)^{-1/2}\right)_{\kappa\kappa^{'}}\varphi_{\kappa^{'}}({\bf r})\,,
$$
or
$$
\psi_{fnlm}({\bf r})=	\varphi_{fnlm}({\bf r})-\frac{1}{2}\sum_{f'n'l'm'}S_{f'n'l'm';fnlm}\,\varphi_{f'n'l'm'}({\bf r}-{\bf R}_{ff'})+
$$
\begin{equation}
\frac{3}{8}\sum_{f'n'l'm'}\sum_{f''n''l''m''}S_{f'n'l'm';f''n''l''m''}S_{f''n''l''m'';fnlm}\,\varphi_{f'n'l'm'}({\bf r}-{\bf R}_{ff'})+\ldots \, .
\end{equation}
Hereafter the orthogonalized functions $\psi_{fnlm}$ can be termed as "quasi-atomic" at variance with nonorthogonalized atomic functions $\varphi_{fnlm}$.

The two-site overlap integrals obey an analog of the Wigner-Eckart theorem\,\cite{var}: 
$$
\int \phi_{l_1m_1}^*({\bf r})\phi_{l_2m_2}({\bf r}-{\bf R}_{12})d{\bf r}=\langle l_1m_1|l_2m_2\rangle =
$$
\begin{equation}
\sum_{kq}(-1)^{l_1-m_1}
\left( \begin{array}{ccc}
l_1 & k & l_2 \\
-m_1 & q & m_2
\end{array} \right)S_k(l_1l_2)\,C^{k}_{q}{}^{*}({\bf R}_{12})\,,	
\end{equation}
where $\left( \begin{array}{ccc}
\cdot & \cdot & \cdot \\
\cdot & \cdot & \cdot
\end{array} \right)$ is a Wigner coefficient\,\cite{var}, $k$ obeys the triangle rule $|l_1-l_2|\leq k \leq l_1+l_2$ given even values of $l_1+ k +l_2 $. In particular, for $l_1=1, l_2=2$ (pd-overlap) $k$ can have only two values: $k=1,3$. In the coordinate system with $O_z$ directed along the radius-vector ${\bf R}_{12}$
$$
\langle 10|20\rangle = S_{pd\sigma}=\,-\,\sqrt{\frac{2}{15}}S_1(pd)+ \sqrt{\frac{3}{35}}S_3(pd)
$$
$$
\langle 1\pm 1|2\pm 1\rangle = S_{pd\pi}=\,-\,\sqrt{\frac{1}{10}}S_1(pd)- \sqrt{\frac{1}{35}}S_3(pd)\,,
$$
where $ S_{pd\sigma}$ and  $ S_{pd\pi}$ are overlap integrals on the   $\sigma $- and $\pi$-bonds, respectively.

Thus, taking into account that the bare atomic functions   $\varphi_{fnlm}({\bf r})$ form a basis of the irreducible representation  $D^{(l)}$ of the rotation group, we can rewrite  $\psi_{fnlm}({\bf r})$ as follows:
$$
\psi_{fnlm}({\bf r})=	\varphi_{fnlm}({\bf r})-\frac{1}{2[l]^{1/2}}\sum_k\sum_{f'n'l'}S_k(f'n'l';fnl)\left[C^k({\bf R}_{ff'})\times\varphi_{f'n'l'}({\bf r}-{\bf R}_{ff'})\right]^l_m
$$
$$
+\frac{3}{8[l]^{1/2}}\sum_{k_1k_2}\sum_{f'n'l'}\sum_{f''n''l''}[l'']^{-1/2}S_{k_1}(f'n'l';f^{\prime\prime}n^{\prime\prime}l^{\prime\prime})S_{k_2}(f^{\prime\prime}n^{\prime\prime}l^{\prime\prime};fnl)
$$
\begin{equation}
\left[C^{k_1}({\bf R}_{f'f^{\prime\prime}})\times [C^{k_2}({\bf R}_{f^{\prime\prime}f})\times\varphi_{f'n'l'}({\bf r}-{\bf R}_{ff'})]^{l^{\prime\prime}}\right]^l_m+... \, ,
\end{equation}
for even values of $(k+l^{\prime}+l),(k_1+k_2+l^{\prime}+l),...$\,. Here we made use of direct products $[\cdot\times\cdot]$ of irreducible tensorial operators\,\cite{var}, $[l]$\,=\,$2l+1$. Unified tensorial form of different terms in the right hand side unambiguously points to the same transformational properties of the bare atomic function $\varphi_{fnlm}({\bf r})$ and its orthogonalized counterpart $\psi_{fnlm}({\bf r})$ (Slater-Koster theorem\,\cite{Slater}), however, with regard to rotations of the laboratory system,
rather than the physical one. Indeed, the former transformation involves coordinates  both
of the electron (${\bf r}$) and the lattice  (${\bf R}_f$) while the latter concerns only the electron coordinates (${\bf r}$). We should emphasize here the preservation of the transformational properties with regard to any rotations rather than certain rotations from the local point group. 

Within a linear approximation the orthogonalization gives rise to a mixing of atomic functions  centered at different sites (a nonlocal hybridization), while within a quadratic approximation one appears the on-site mixing of different atomic $nlm$-functions (a local hybridization). The quadratic correction can be written as follows:  
\begin{equation}
\Delta\psi_{fnlm}({\bf r})=\sum_{k}\sum_{f''}\sum_{n'l'}S_{k}(ff^{\prime\prime}n'l';ff^{\prime\prime}nl)\left[C^k({\bf R}_{ff^{\prime\prime}})\times\varphi_{fn'l'}({\bf r})\right]^l_m	\, ,
\end{equation}
where
$$
S_{k}(ff^{\prime\prime}n'l';ff^{\prime\prime}nl)=\frac{3}{8[l]^{1/2}}\sum_{k_1k_2}\sum_{n''l''}(-1)^{k_2}[k]^{1/2}\left[ \begin{array}{ccc}
k_1 & k_2 & k \\
0 & 0 & 0
\end{array} \right]\left\{ \begin{array}{ccc}
k_1 & k_2 & k \\
l^{\prime} & l & l^{\prime\prime}
\end{array} \right\}
$$
\begin{equation}
S_{k_1}(fn'l';f^{\prime\prime}n^{\prime\prime}l^{\prime\prime})S_{k_2}(f^{\prime\prime}n^{\prime\prime}l^{\prime\prime};fnl)
\label{2}	
\end{equation}
given even values of $k+l^{\prime}+l$, $\left[ \begin{array}{ccc}
k_1 & k_2 & k \\
0 & 0 & 0
\end{array} \right]$ and $\left\{ \begin{array}{ccc}
k_1 & k_2 & k \\
l^{\prime} & l & l^{\prime\prime}
\end{array} \right\}$ are the Clebsch-Gordan coefficient and $6j$-symbol, respectively\,\cite{var}. Structure of $S_{k}(ff^{\prime\prime}n'l';ff^{\prime\prime}nl)$, in particular, summation on $n^{\prime\prime}l^{\prime\prime}$ implies significant troubles with its reliable estimates. Anyhow this quantity is on the order of $S_{pd}^2\sim$\,0.01 for typical 3d oxides.
It is worth noting that the orthogonalization does enhance the role of high-energy unfilled excited states which are characterized by a strong overlap with atomic functions of the filled orbitals of neighboring atoms.

Application of the irreducible tensor technique to the orthogonalized $fnlm$-functions  provides a regular procedure for a revision of different effects which are typical for many-electron atomic systems.
Below we address different properties of quasi-atoms which are composed of "orthogonalized" electrons. 
%Ниже мы рассмотрим различные свойства квазиатомов, сформированных из квазиатомных электронов, а также их взаимодействие.
%Переход к ортогонализованным электронным состояниям позволяет проиллюстрировать в рамках регулярной процедуры роль различных эффектов, типичных для многоатомных систем.  

%\subsection{Особенности "`свободных"' квазиатомов}

\section{"Delocalization"  of the orbital atomic operators}
Delocalization of atomic orbitals in orthogonalized $\psi_{fnlm}$ functions gives rise to an unconventional effect of a "delocalization"  of the orbital atomic operators. 
Indeed, a local irreducible tensorial operator 
 ${\hat V}^a_{\alpha}(fnl)$, working within the conventional nonorthogonalized basis set of
 the $(2l+1)$ $nlm$-states for the $f$-atom according to the Wigner-Eckart theorem
\begin{equation}
\langle nlm|{\hat V}^a_{\alpha}(fnl)|nlm^{\prime}\rangle = (-1)^{l-m} \left( \begin{array}{ccc}
l & a & l \\
-m & \alpha & m^{\prime}
\end{array} \right)\langle nl||{\hat V}^a(fnl)||nl\rangle \, ,	
\end{equation}
will be described within the basis set of the orthogonalized counterparts $\psi_{fnlm}$ by an equivalent operator 
$$
{\hat V}^a_{\alpha}(fnl)+2\sum_{kb}(-1)^b[l,a]^{1/2}S_{k}(ff^{\prime\prime}nl;ff^{\prime\prime}nl)
\left\{\begin{array}{ccc}
k & a & b \\
l & l & l
\end{array}\right\}
\left[{\hat V}^b(fnl)\times C^k({\bf R}_{ff''})\right]^a_{\alpha}
$$
$$
+\frac{1}{4[a]^{1/2}}\sum_{k_1k_2kb}\sum_{f'n'l'}[b][k]^{1/2}S_{k_1}(f'n'l';fnl)S_{k_2}(f'n'l';fnl)\langle n^{\prime}l^{\prime}||{\hat V}^a(f'n'l')||n^{\prime}l^{\prime}\rangle
$$
\begin{equation}
\left[ \begin{array}{ccc}
k_1 & k_2 & k \\
0 & 0 & 0
\end{array} \right]\left\{ \begin{array}{ccc}
k_1 & l^{\prime} & l \\
k_2 & l^{\prime} & l \\
k & a & b
\end{array} \right\}
\left[{\hat v}^b(fnl)\times C^k({\bf R}_{ff'})\right]^a_{\alpha}
\label{V}
\end{equation}
given even values of $k$ and $(a+b)$, $\left\{ \begin{array}{ccc}
k_1 & l^{\prime} & l \\
k_2 & l^{\prime} & l \\
k & a & b
\end{array} \right\}$ is the $9j$-symbol\,\cite{var}, $[a,l]$\,=\,$(2a+1)(2l+1)$. Here ${\hat v}^b(fnl)$ is a $b$-rank irreducible tensorial operator whose submatrix element  equals one. The Exp.\,(\ref{V}) describes the delocalization effect for different quantities related to tensorial operators. It is worth noting the emergence of tensorial operators  ${\hat v}^b_{\beta}(fnl)$  whose rank differs from that of ($a$) the bare operator.

\section{Effective orbital moment}

Obviously that at variance with the $\varphi_{fnlm}({\bf r})$ functions the $\psi_{fnlm}({\bf r})$ functions are not the eigenfunctions for operators ${\hat {\bf l}}^2$ and ${\hat l}_z$ that is for the square and  $z$-component of the orbital momentum operator, though they form a basis of the irreducible representation $D^{(l)}$ of the rotation group. Nonetheless, one can introduce an effective orbital momentum,  or quasi-momentum ${\hat {\tilde{\bf l}}}$ as follows:
\begin{equation}
{\hat {\tilde{\bf l}}}=-i\hbar\left([{\bf r}\times{\bf \nabla}_r]+\sum_{{\bf R}_f}[{\bf R}_f\times{\bf \nabla}_{R_f}]\right)	\, .
\end{equation}
By analogy with a free atom we can introduce an addition of the orbital quasi-momenta thus forming many-electron configurations and wave functions $|{\tilde L}SM_{{\tilde L}}M_S\rangle$ for the ${}^{2S+1}{\tilde L}$  terms of the many-electron quasi-free atom which incorporates all the effects of the orthogonalization of one-electron states for different sites. 

The orthogonalization procedure preserves many though not all the advantages of the free atom theory based on the application of the theory of the rotation group. Thus the Wigner-Eckart theorem for conventional irreducible tensorial operators such as spherical harmonic $C^k_q({\bf r})$, orbital momentum $\hat {\bf l}$ and other orbital operators acting in the conventional ${\bf r}$ space does not work on $\psi_{fnlm}({\bf r})$-functions because of matrix elements will depend parametrically on the lattice vectors. 

However, in practice, e.g., for Zeeman coupling one should address a true orbital momentum whose relations with effective orbital momentum  will have a nontrivial form. Within a  $|\tilde{L}M_{\tilde{L}}\rangle$ multiplet these relations are as follows: 
\begin{equation}
	\hat L_i= a_{ij}\tilde{L}_j+a_{ijkl}\tilde{L}_j\tilde{L}_k\tilde{L}_l+...\,   ,
\end{equation}
where we meet with unconventional tensorial linear and different nonlinear terms.

The second rank tensor $a_{ij}$ for the systems such as a 3d ion in oxides can be written as follows:
\begin{equation}
	a_{ij}=a\delta_{ij}+\Delta\,a_{ij}\, \
\end{equation}
where $a\leq 1$, but $\Delta\,a_{ij}\sim S^2$ ($S$ is a cation-anion overlap integral), 
and the difference between $a$ and one is on the order of  $S^2$. Point symmetry puts distinct limitations on the  $a_{ij}$ tensor, e.g., for a cubic symmetry  $\Delta\,a_{ij}=0$. 

Complex nonlinear relations between true and effective orbital momenta results in a nontrivial form of Zeeman coupling 
\begin{equation}
	\hat V_{Z}=\mu_B ({\hat {\tilde{\bf L}}}\cdot{\bf H})=\mu_B\left\{{\hat {\tilde{\bf L}}}{\hat a^{(2)}}{\bf H}+{\hat {\tilde{\bf L}}}{\hat {\tilde{\bf L}}}{\hat {\tilde{\bf L}}}{\hat a^{(4)}}{\bf H}+...\right\}\,  ,
\end{equation}
where we make use of a symbolic form of a tensorial product of vector operators. Thus the Zeeman coupling acquires novel features due to the orthogonalization procedure: i) a reduction of the effective orbital $g$-factor as compared with its bare value $g_l$\,=\,1 with its anisotropy for low-symmetry sites; ii) emergence of a nonlinear Zeeman coupling. These features evolve from the orthogonalization procedure as a zero order perturbation effect prior to effects of crystal field and covalency.

\section{Electric dipole moment of a "free quasi-atom"}

Orthogonalized  $\psi_{fnlm}({\bf r})$ functions for noncentrosymmetric sites $f$, at variance with bare $\varphi_{fnlm}({\bf r})$ functions, do not have a definite parity. This immediately gives rise to a nonzero electric dipole polarization of such a "free quasi-atom".    

Taking into account the local hybridization effects (\ref{2}) we can write one-electron matrix element of electric dipole moment as follows
$$
\langle \psi_{fnlm}|{\hat d}_q|\psi_{fnlm^{\prime}}\rangle =\frac{2e}{\sqrt{3}}[l]^{\frac{1}{2}}\sum_{a,k}\sum_{f''}\sum_{n'l'}[a](l||C^1||l^{\prime})r_{nl;n'l'}S_{k}(ff^{\prime\prime}n'l';ff^{\prime\prime}nl)
$$
\begin{equation}
\left\{ \begin{array}{ccc}
1 & k & a \\
l & l & l^{\prime}
\end{array} \right\}
(-1)^{l-m}
\left( \begin{array}{ccc}
l & a & l \\
-m & \alpha & m^{\prime}
\end{array} \right)\left[ \begin{array}{ccc}
a & k & 1 \\
\alpha & q^{\prime} & q
\end{array} \right]^{}_{}
C^{k}_{q^{\prime}}({\bf R}_{ff^{\prime\prime}})\, ,
\end{equation}
 where  $a=0,2,...2l $ and $k=a\pm 1$ are even and odd numbers, respectively. Here $(l||C^1||l^{\prime})$ is a submatrix element of a tensorial harmonic\,\cite{var},
 $$
 r_{nl;n'l'}=\int_{0}^{\infty}R_{nl}(r)r^3R_{n^{\prime}l^{\prime}}(r)dr
 $$
 is a dipole radial integral. 
Thus within the $\psi_{fnlm}$ basis set with a certain $fnl$ the dipole moment operator can be replaced by an effective (equivalent) operator as follows  
\begin{equation}
	{\hat d_q}= \sum_{a,k,f^{\prime}}d_{ak}(R_{ff^{\prime}})\left[{\hat v}^a({\bf l})\times C^k({\bf R}_{ff^{\prime}})\right]^{1}_{q}\, ,
	\label{dq}
\end{equation}
where ${\hat v}^{a}_{\alpha}(l)$ is an orbital $a$-rank irreducible tensorial operator with the unit submatrix.
In Cartesian coordinates the Exp.\,(\ref{dq}) can be written as follows
\begin{equation}
	{\hat d}_i=d_i^{(0)}+\frac{1}{2}d_{ijk}^{(2)}\{{\hat l}_j,{\hat l}_k\}+\mbox{terms with a=4,...}\, ,
\end{equation}
where $\{{\hat l}_j,{\hat l}_k\}={\hat l}_j{\hat l}_k+{\hat l}_k{\hat l}_j$.
Of a particular interest the terms with nonzero values $a$\,=\,2,\,4,..., which directly relate the electric polarization to the degenerated or quasi-degenerated orbital state, in particular, to quadrupole ($a$\,=\,2), octupole  ($a$\,=\,4) electronic momenta of quasi-atoms and give rise to an orbital magnetoelectric coupling. Strictly speaking, the electric polarization induced by the orthogonalization can amount to big values due to both big overlap  and radial integrals $r_{nl;n'l'}\sim 1\,\AA$.

The expression (\ref{dq}) can be easily generalized for many-electron quasi-atomic configurations. For certain terms ${}^{2S+1}L$ for the $nl^N$ shell of equivalent electrons one should replace the one-electron operator ${\hat v}^{a}_{\alpha}({\bf l})$ in (\ref{dq}) by  the equivalent many-electron orbital operator ${\hat V}^{a}_{\alpha}({\bf L})$ which acts on the $\left|LM_L\right\rangle$ basis set according to the Wigner-Eckart theorem
\begin{equation}
\langle LM_L|{\hat V}^a_{\alpha}({\bf L}))|L^{\prime}M_{L^{\prime}}\rangle = (-1)^{L-M_L} \left( \begin{array}{ccc}
L & a & L^{\prime} \\
-M_L & \alpha & M_{L^{\prime}}
\end{array} \right) U^{(a)}_{SL;SL^{\prime}} \, ,	
\end{equation}
where $U^{(a)}_{SL;SL^{\prime}}$ is the spectroscopic Racah coefficient\,\cite{Sobelman}.
For the ${}^{2S+1}L_J$ multiplets the ${\hat V}^{a}_{\alpha}({\bf l})$ operator in (\ref{dq}) should be replaced by the equivalent many-electron orbital operator ${\hat V}^{a}_{\alpha}({\bf J})$ which acts on the $\left|SLJM_J\right\rangle$ basis set according to the Wigner-Eckart theorem
\begin{equation}
\langle SLJM_J|{\hat V}^a_{\alpha}({\bf J}))|SL^{\prime}J^{\prime}M_{J^{\prime}}\rangle = (-1)^{J-M_J} \left( \begin{array}{ccc}
J & a & J^{\prime} \\
-M_J & \alpha & M_{J^{\prime}}
\end{array} \right) U^{(a)}_{SLJ;SL^{\prime}J^{\prime}} \, ,	
\end{equation} 
where\,\cite{Sobelman}
\begin{equation} 
 	U^{(a)}_{SLJ;SL^{\prime}J^{\prime}}= (-1)^{S+a+L'+J} \sqrt{\left(2J+1\right)\left(2J'+1\right)} 
 	\left\{
 	{\arraycolsep=0.2em
 	\begin{array}{ccc} 
 		{L} & {J} & {S} 
 		\\ 
 		{J'} & {L'} & {a} 
 	\end{array}
 	} 	
 	\right\}
 	U^{(a)}_{SL;SL^{\prime}}\,.
\end{equation}

Consideration of the linear in overlap effects needs in a knowledge of two-site dipole matrix elements which can be written as follows:
$$ 
\langle \varphi_{fnlm}|{\hat d}_q|\varphi_{f^{\prime}n^{\prime}l^{\prime}m^{\prime}}\rangle = 
\sum_{a,\alpha ,k,q^{\prime}}(-1)^{l-m}
\left( \begin{array}{ccc}
l & a & l^{\prime} \\
-m & \alpha & m^{\prime}
\end{array} \right) 
$$
\begin{equation}
\langle fnl||d^{ak}||f^{\prime}n^{\prime}l^{\prime}\rangle 
\left[ \begin{array}{ccc}
a & k & 1 \\
\alpha & q^{\prime} & q
\end{array} \right]
C^{k}_{q^{\prime}}({\bf R}_{ff^{\prime}})\, .
\end{equation} 
Accordingly, the contribution to  dipole matrix on the $\psi_{fnlm}$  basis with a certain set of quantum numbers $fnl$ acquires a tensorial form which is similar to (\ref{dq}). As an obvious practical implication of the Exp. (\ref{dq}) we should point to the calculation  of the probabilities of the intra-configurational electro-dipole transitions, e.g., d-d transitions for 3d compounds or f-f transitions for 4f compounds, which are dipole-allowed for noncentrosymmetric quasi-atoms.

\subsubsection{Judd-Ofelt theory of effective electric dipole moment for noncentrosymmetric ions}

The Judd-Ofelt theory of effective electric dipole moment for noncentrosymmetric ions\,\cite{Judd} takes into account the admixing of configurations of opposite parity
due to the odd-parity crystal field
\begin{equation}
	\hat V_{cf}=\sum_{kq}A_{kq}^*r^kC_q^k(\theta ,\phi ) \, ,
\end{equation}
where $A_{kq}$ are crystal field parameters. The effective dipole moment operator can be written out similarly to (\ref{dq}) as follows 
\begin{equation}
	{\hat d_q}= \sum_{a,k}\tilde{d}_{ak}\left[{\hat v}^a({\bf l})\times A^k\right]^{1}_{q}\, ,
	\label{dqq}
\end{equation}
where
\begin{equation}
\tilde{d}_{ak}=-\frac{1}{\sqrt{3}}(2a+1)\Xi (ka)\, ,	
\end{equation}
and
$$
\Xi (ka)=2(2a +1)\sum_{l^{\prime}}(2l^{\prime}+1)(-1)^{l+l^{\prime}}
\left\{ \begin{array}{ccc}
1 & a & k \\
l & l^{\prime} & l
\end{array} \right\}
$$
\begin{equation}
\left( \begin{array}{ccc}
l & 1 & l^{\prime} \\
0 & 0 & 0
\end{array} \right)
\left( \begin{array}{ccc}
l^{\prime} & k & l \\
0 & 0 & 0
\end{array} \right)\frac{\langle 4f|r|n^{\prime}l^{\prime}\rangle \langle 4f|r^k|n^{\prime}l^{\prime}\rangle}{E_{n^{\prime}l^{\prime}}-E_{nl}}
\,  .
\end{equation}

\section{The overlap contribution to orbital magnetoelectric coupling}
For paramagnetic ions with an orbital (quasi)degeneracy the thermal expectation value 
$\left\langle {\hat V}^{a}_{\alpha}(l)\right\rangle$ in (\ref{dq}) can strongly depend on the magnetic field, either internal molecular or the external one thus providing an effective magnetic control of electric polarization. Indeed, in the absence of crystal field and spin-orbital effects the thermal expectation value  can be represented as follows:
\begin{eqnarray}
 \left< {\hat V}^{a}_{\alpha}(l)\right>_{T} = \left< {\hat V}^{a}_{0}(l)\right>_{T}
   C_{\alpha}^{a}({\bf {L}}),  \,\,\,
 \left< V_{0}^{a}\right>_T = \left< V_{0}^{a}\right>_{0}
  \rho_{a}(H,T), \label{eq:ave}
\end{eqnarray}
where $C_{\alpha}^{a}({\bf {L}})$   is tensorial spherical harmonic with a classical vector ${\bf {L}}$ as an argument,
\(\left< V_{\alpha}^{a}\right> \) the thermal expectation value with
\(
 \left< V_{0}^{a}\right>_{0} =
 \left( \begin{array}{ccc}
L & a & L \\
-L & 0 & L
\end{array} \right),
\) and \(\rho_{a}(H,T)\) is a temperature factor, e.g.:
\begin{eqnarray}
 \rho_{0} = 1 \quad ;
 \rho_{2} = \frac{\left< 3L_{z}^{2}-L(L+1)\right>}{L(2L-1)}.
\end{eqnarray} 
Within a molecular field approximation $C_{\alpha}^{a}({\bf {L}})\equiv C_{\alpha}^{a}({\bf {H}})$. 
Thus in frames of our simplifications we arrive at a very interesting expression 
\begin{equation}
	\left<{\hat d_q}\right>= \sum_{a,k,f^{\prime}}d_{ak}(R_{ff^{\prime}})\left< V_{0}^{a}(L)\right>_{0}
  \rho_{a}(H,T)\left[C^{a}({\bf {H}})\times C^k({\bf R}_{ff^{\prime}})\right]^{1}_{q}\, ,
  \label{dL}
\end{equation}
that demonstrates a magnetic control of the electric polarization in the most clear way.
In Cartesian coordinates the Exp.\,(\ref{dq}) can be written as follows
\begin{equation}
	{\hat d}_i=d_i^{(0)}+d_{ijk}^{(2)}(H,T)h_j\,h_k+\mbox{terms with a=4,...}\, ,
\end{equation}
where $d_{ijk}^{(2)}(H,T)\propto\rho_{2}(H,T)$ and ${\bf h}$\,=\,${\bf H}/H$.

For rare-earth ions with 4f$^n$ shell and strong spin-orbital coupling the electric polarization for the $SLJ$-multiplet can be easily written as follows:
$$
	\left<{\hat d_q}\right>= \sum_{a,k,f^{\prime}}(-1)^{S+a+L+J} [J] 
 	\left\{
 	{\arraycolsep=0.2em
 	\begin{array}{ccc} 
 		{L} & {J} & {S} 
 		\\ 
 		{J} & {L} & {k} 
 	\end{array}
 	} 	
 	\right\}
$$
\begin{equation} 	
 	d_{ak}(R_{ff^{\prime}})\left< V_{0}^{a}(J)\right>_{0}
  \rho_{a}(T)\left[C^{a}({\bf {H}})\times C^k({\bf R}_{ff^{\prime}})\right]^{1}_{q}\, .
  \label{dJ}
\end{equation}
It should be noted that the expressions (\ref{dL}) and (\ref{dJ}) define both the field and temperature dependence of the electric dipole moment, however, given a complete neglect of the the crystal field quenching effects. These effects are usually much stronger than Zeeman coupling, in particular, for 3d ions. However, for some rare earth ions with so-called  quasi-doublets in ground state the Zeeman and crystal field effects can compete with each other, and our Exps. (\ref{dq}) and (\ref{dJ}) predict a rather strong magnetoelectric coupling. Such a situation is realized, e.g., for Tb$^{3+}$ ion in the multiferroic terbium manganate TbMn$_2$O$_5$ (see, e.g., Ref.\,\cite{Chang}).

For the 3d ions the crystal field quenches both the orbital momenta and orbital magnetoelectric coupling. However, due to the spin-orbital coupling/mixing the orbital effects give rise to different effective spin interactions, e.g., the emergence of an orbital contribution to effective spin $g$-tensor and single-ion spin anisotropy. Furthermore, the orbital operator  
${\hat V}^{a}_{\alpha}({\bf L})$ given $a$\,=\,2, 4, ... for orbitally nondegenerated ground state can be replaced by an effective spin operator. This is  relatively easy to perform for so-called S-type 3d ions which have an orbitally nondegenerated ground state of the $A_{1}$ or $A_{2}$ type for a high-symmetry octahedral, cubic or tetrahedral crystal field. These are ions, e.g.,  with 3d$^3$ (Cr$^{3+}$, Mn$^{4+}$), 3d$^5$ (Fe$^{3+}$, Mn$^{2+}$), 3d$^8$ (Ni$^{2+}$, Cu$^{3+}$) configurations with an octahedral surroundings. Strictly speaking, first we should replace the 
${\hat V}^{a}_{\alpha}({\bf L})$ by a linear combination
\begin{equation}
{\hat V}^{a\gamma}_{\nu}({\bf L})=\sum_{\alpha} c_{a\alpha}^{\gamma\nu}	{\hat V}^{a}_{\alpha}({\bf L})\, ,
\label{V(L)}
\end{equation}
which form a basis of the irreducuble representation $D^{\gamma}$ (line $\nu$) of the point symmetry group, $O_h$ or $T_d$. Here $\gamma$\,=\,$E$ or $T_2$ given $a$\,=\,2, $\gamma$\,=\,$A_1$, $E$, $T_1$ or $T_2$ given $a$\,=\,4. Next we can replace the orbital operator by a spin equivalent as follows
\begin{equation}
\left<{\hat V}^{a\gamma}_{\nu}({\bf L})\right>_{GS}=\eta_{a\gamma}{\hat V}^{a\gamma}_{\nu}({\bf S})\, ,	
\end{equation}
where the $\left<...\right>_{GS}$ means a mapping on the ground state. The $\eta_{a\gamma}$ parameters depend on the type of the 3d ion:$\eta_{2\gamma}\sim\left(\zeta_{3d}/\Delta\right)^2\leq 10^{-3}$, here $\zeta_{3d}$ is a spin-orbital constant, $\Delta$ is a mean energy of excited $T$-terms.  
Thus for S-type ions we can transform the orbital dipole moment operator (\ref{dq}) into an effective spin operator which acts within the $SM_S$-multiplet of the ground state as follows
\begin{equation}
	{\hat d}_q= \sum_{a\gamma\nu}R^{(a\gamma )}_{q\nu}{\hat V}^{a\gamma}_{\nu}({\bf S})\, .
	\label{dS}
\end{equation}
where
\begin{equation}
R^{(a\gamma )}_{q\nu}=\sum_{a,k,f^{\prime}}d_{ak}(R_{ff^{\prime}})\eta_{a\gamma}\sum_{\alpha q^{\prime}}c_{a\alpha}^{\gamma\nu *}\left[ \begin{array}{ccc}
a & k & 1 \\
\alpha & q^{\prime} & q
\end{array} \right]
C^{k}_{q^{\prime}}({\bf R}_{ff^{\prime}})\, .
\end{equation}
As in (\ref{V(L)}) ${\hat V}^{a\gamma}_{\nu}({\bf S})=\sum_{\alpha} c_{a\alpha}^{\gamma\nu}	{\hat V}^{a}_{\alpha}({\bf S})$, however, with a more simple spin matrix for ${\hat V}^{a}_{\alpha}({\bf S})$:
\begin{equation}
\langle SM_S|{\hat V}^a_{\alpha}({\bf S}))|SM_{S^{\prime}}\rangle = (-1)^{S-M_S} \left( \begin{array}{ccc}
S & a & S \\
-M_S & \alpha & M_{S^{\prime}}
\end{array} \right) \, .	
\end{equation}
The spin irreducible tensorial operators ${\hat V}^{a\gamma}_{\nu}({\bf S})$ can be transformed into a Cartesian form, e.g.
\begin{equation}
{\hat V}^{2E}_{0}({\bf S})=2\,\frac{3{\hat S}_z^2-S(S+1)}{\sqrt{(2S+3)^{(5)}}};\,{\hat V}^{2E}_{2}({\bf S})=2\sqrt{3}\frac{({\hat S}_x^2-{\hat S}_y^2)}{\sqrt{(2S+3)^{(5)}}}\,, ...\,,	
\end{equation}
where $(2S+3)^{(5)}=(2S+3)(2S+2)...(2S-1)$.
In Cartesian coordinates the Exp.\,(\ref{dS}) can be reduced to a standard form
\begin{equation}
	{\hat d}_{i}=d_i^{(0)}-\frac{1}{2}R_{ijk}\{{\hat S}_j,{\hat S}_k\}+\mbox{terms with a=4,...}\, ,
	\label{R}
\end{equation}
widely adopted in the theory of the electric field effects in electron spin resonance (ESR)\,\cite{Mims}.

Recently the magnetoelectric effect due to local noncentrosymmetry was addressed in Ref.\,\cite{Ter-Oganessian}. The authors started with the nonorthogonalized basis set of 3d orbitals, next  considered an odd-parity crystal field and spin-orbital coupling as perturbations. Finally they arrived at an expression which is similar to (\ref{R}). %$d_{\alpha}=-R_{\alpha ij}S_iS_j$ which was used for studying a microscopic nature of the magnetoelectricity in several multiferroics. We rewrite their Exps.\,(5) in the form, widely adopted in the theory of the electric field effects in electron spin resonance (ESR)\,\cite{Mims}. 
Furthermore, in fact, the microscopic consideration in Ref.\,\cite{Ter-Oganessian} reproduces, however, with strong simplification the well-known paper by Kiel and Mims\,\cite{Kiel-CaWO} on electric field effect in ESR for Mn$^{2+}$ ions in CaWO$_4$. 

Interestingly that the experimental findings of this and many other papers on the electric field effect in ESR\,\cite{Mims} can be used for a direct estimation of the single-ion contribution to the magnetoelectric coupling in different multiferroics. For instance, for well known multiferroic MnWO$_4$\,\cite{MWO} one might use the parameters $R_{ijk}$ measured for Mn$^{2+}$ ions in CaWO$_4$\,\cite{Kiel-CaWO}:being normalized to unit cell volume of MnWO$_4$\,\cite{MWO} ($V_c$\,$\approx$\,138\,$\AA^3$) these are as follows: $|R_{123}|$\,$\approx$\,1.8, $|R_{113}|$\,$\approx$\,1.5, $|R_{311}|$\,$\approx$\,0.3, $|R_{312}|$\,$\approx$\,5.0\,$\mu$C\,m$^{-2}$ ($R_{223}$\,=\,-$R_{113}$, $R_{213}$\,=\,$R_{123}$). It should be noted that for Mn$^{2+}$ ions in SrWO$_4$ these parameters are nearly two times larger\,\cite{Kiel-SrWO}. Taking account  of two Mn$^{2+}$ ions in unit cell and a rather large value ($\leq$\,6) of the spin factor in (\ref{R})  one may conclude that the single-ion mechanism can be a significant contributor to the ferroelectric polarization observed in MnWO$_4$ ($P_b\sim$\,50\,$\mu$C\,m$^{-2}$)\,\cite{MWO}.
Interestingly that very recent quantitative estimates of the spin-dependent ferroelectric polarization in MnWO$_4$ based on the low-energy model, derived
from the first-principles electronic structure calculations\,\cite{Solovyov} showed values which are an order of magnitude less than  the experimental ones.
      
It is worth noting that the single-ion term (\ref{R}) does not produce the magnetoelectric coupling for quantum spins $S=\frac{1}{2}$ (e.g., Cu$^{2+}$) due to a kinematic constraint:$0\leq a\leq 2S$.

 \section{Conclusion}
 
Making use of the L\"{o}wdin orthogonalization for one-electron atomic $\varphi_{fnlm}({\bf r})$ wave functions we arrive at a basis set of orthogonalized $\psi_{fnlm}({\bf r})$ orbitals which formally preserve symmetric properties of the bare $\varphi_{fnlm}({\bf r})$ orbitals. Instead of many-electron atomic configurations composed of non-orthogonalized $nlm$-orbitals we arrive at quasi-atoms composed of the orthogonalized $nlm$-counterparts. Formal conservation of the  rotational symmetry allows to apply the powerful technique of irreducible tensorial operators\,\cite{var} and Racah algebra\,\cite{Sobelman} to description of the quasi-atoms. As an illustration we addressed a single-ion contribution to magnetoelectric coupling which is usually missed in current studies on multiferroics. A regular procedure has been developed 
for calculation of the overlap contribution to the single-ion orbital magnetoelectric coupling
both for 3d- and 4f-ions. In a sense the overlap contribution resembles the point charge contribution to a crystal field, this correctly describes both the lattice symmetry and  
the symmetry of electronic states, and provides reasonable semiquantitative estimates. Furthermore, making use of experimental data for electric field effect in ESR we have shown that single-ion magnetoelectric coupling can be a leading mechanism of multiferroicity, e.g., in MnWO$_4$.

\ack
My thanks to A.I. Liechtenstein, Yu.D. Panov and N.V. Ter-Oganessian for useful discussions.
  The work is supported by the Ural Federal University (No.211 Decree of the Government of RF of 16th March 2013) and the  RFBR grant No. 12-02-01039.

\end{document}